\documentclass[12pt]{article}

\usepackage{amsmath}
\usepackage{amsfonts}
\usepackage{amssymb}
\usepackage[small]{caption}
\usepackage{color}
\usepackage{fullpage,citesort,epsfig,graphics,amsbsy}
\usepackage{cancel}
\usepackage{psfrag}
\usepackage{verbatim}
\newcommand{\bea}{\begin{eqnarray}}
\newcommand{\eea}{\end{eqnarray}}

\newcommand{\nn}{\nonumber}
\newcommand{\ee}[2]{\epsilon_{#1} \!\cdot\! \epsilon_{#2}}

\long\def\symbolfootnote[#1]#2{\begingroup
\def\thefootnote{\fnsymbol{footnote}}\footnote[#1]{#2}\endgroup} 

\begin{document}
\begin{titlepage}
\begin{flushright}
\phantom{UFIFT-HEP-10-}
\end{flushright}

\vskip 2.5cm

\begin{center}
\begin{Large}
{\bf A Note on High Energy Scattering of Open Superstrings\footnote{Supported 
in part by the Department
of Energy under Grant No. DE-FG02-97ER-41029}
}
\end{Large}

\vskip 2cm
{\large 
Francisco Rojas\,\symbolfootnote[1]
{E-mail  address: {\tt frojasf@phys.ufl.edu}}
}
\vskip0.20cm
{\it Institute for Fundamental Theory\\
Department of Physics, University of Florida,
Gainesville FL 32611}
\vskip 1.0cm
\end{center}

\begin{abstract}
\hspace{-20pt}  
We study the Regge and hard scattering limits of the one-loop amplitude for massless open string 
states in the type I theory in flat space. For hard scattering we find the exact kinematic dependence 
in terms of the scattering angle of the factor multiplying 
the known exponential falloff, without relying on a saddle point 
approximation for the integration over the cross ratio. This bypasses the issues of estimating the 
contributions from flat directions as well as those that arise from fluctuations of the gaussian 
integration about the saddle point. This result allows for a straightforward computation of the 
small-angle behavior of the hard scattering regime and we find complete agreement with the Regge limit at 
high momentum transfer, as expected.
\end{abstract}
\vfill
\end{titlepage}
\section{Introduction}
Open string amplitudes including one-loop corrections in the high energy regime have been studied 
since the very early days of string theory \cite{Neveu:1970iq,AAM,OPE}, and subsequently  the subject was 
re-taken by 
many other authors in the 1980's 
\cite{Clavelli,GrossManes}. In \cite{GrossManes} the authors analyzed the (2+2) non-planar amplitude at one-loop, 
i.e., the annulus diagram with two 
external states attached to each boundary, and found the (now well-known) characteristic exponential 
falloff of stringy amplitudes in this regime. The coefficient multiplying this exponential behavior, 
which contains dependence on the scattering angle, involves the typical problem of inversion in the 
theory of elliptic modular functions. As a consequence of this, the angular dependence in the 
aforementioned coefficient could only be expressed in terms of an infinite series.

For the case of the planar and non-orientable amplitudes, Gross and Ma\~nes \cite{GrossManes} 
studied this high energy regime a fixed scattering angle and found that, contrary to the (2+2) non-
planar case, they do not possess a dominant saddle point in the interior of the integration region. 
Moreover, they were able to show that the dominant contributions come from the boundaries of this 
region, the one where the annulus shrinks to a point being the dominant boundary in this case. 

The study of the fixed-angle limit of the one-loop amplitude in different situations has been carried out 
by many authors \cite{GrossManes,Bachas,Barbon,DiVecchia}, but as far as we aware of, we believe 
that the exact dependence on the scattering angle for the amplitude we study here has not been 
worked out in the literature in a closed form. 

We organize this short note as follows: In section 2 we review the calculation of the Regge limit of 
the sum of the planar and non-orientable diagrams of the type I theory. We also compute its 
large momentum transfer limit ($|t|\to \infty$) in order to make a comparison with the small-angle 
behavior of the hard scattering limit which we also review in this section. In section 3, by making use 
of an identity originally used in \cite{GS}, although in a different context, we compute the exact form of the coefficient that multiplies the 
exponential falloff of the amplitude in the high-energy regime at fixed scattering angle. This permits a staightforward evaluation of the the 
hard scattering amplitude in the limit where $t \ll s$, which indeed matches with the Regge behavior 
at high momentum transfer computed in section 2.

\section{High-energy scattering of the type I open superstrings}
\subsection{Regge behavior at one-loop}
We begin by computing the Regge limit, i.e. we take $s\to -\infty$ with $t$ held fixed of the one-loop 
amplitude for type I open superstrings. The details of the calculation are basically the same as the ones 
computed for the type 0 string in \cite{RojasThorn} with the only difference being the nature of the 
cancellation of divergences due to the propagation of closed string tachyons and dilatons. In 
\cite{RojasThorn} the remnants of closed string tachyon divergences were cancelled by the inclusion of a 
counterterm which, after analytic continuation using a momentum regulator, 
turned out to be zero in the Regge limit. The ``would-be'' subleading divergences due to closed string 
dilatons were simply absent with the inclusion of D$p$-branes as long as $p<7$.

\begin{figure}[ht!]
\centering
\includegraphics[scale=0.50]{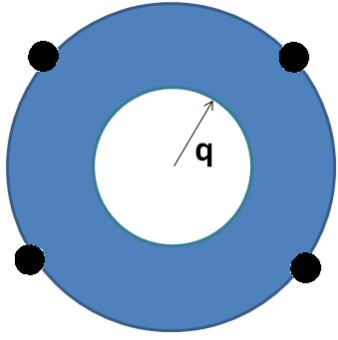}
\caption{For the planar one-loop amplitude all the external states lie at only one of the two boundaries, 
and the integration over $q$ is represented as a radial variable. The region $q\sim 0$ corresponds to highly energetic open strings 
and it gives the dominant contribution in the hard scattering regime.}
\label{fig:planarloop}
\end{figure}

The amplitude for four massless vector states is much simpler in the superstring compared to the type 0 
model, because in the former the full polarization structure can be factored out of the loop integration, 
whereas in the latter each combination of polarization vectors must be worked out separately. For the 
$SO(32)$ gauge group the planar and non-orientable one-loop diagrams combine to give a finite expression \cite{Green:1984ed} and we focus our attention on this case in this article. The amplitude for each diagram (planar and non-orientable) was computed long time ago (see for instance \cite{GSW2}) and for the $SO(32)$ gauge group they can be combined as
\bea
A_P+A_N= 16\pi^3 g^4  G_P K \int_0^1 \frac{dq}{q} \left[F(q^2)-F(-q^2)\right]
\label{full}
\eea
with
\bea
F(q^2) &=&   \int_R\prod_{i=1}^3  d\theta_i\ \prod_{i<j}\psi(\theta_{ji})^{2\alpha' k_i \cdot k_j}\nn\\
\psi(\theta) &=& \sin \theta \prod_{n=1}^{\infty} \frac{1-2q^{2n}\cos2\theta+q^{4n}}{(1-q^{2n})^2} 
\label{fulldef}
\eea
and $K$ is the kinematic factor which can be found, for example in \cite{GSW2}. The region of integration $R$ is given by $0 < \theta_2 < \theta_3 < \theta_4 < \pi$, $\theta_{ji}\equiv\theta_j-\theta_i$, and $G_P$ is the group theory factor $ G_P={\rm Tr}(\lambda_1\lambda_2\lambda_3\lambda_4) $.
We can now go ahead and compute the behavior of this expression for $s\to -\infty$ holding $t$ fixed. In this limit, the amplitude is dominated by the region $\theta_2\sim \theta_3$ and $\theta_4 \sim \pi$. Writing
\bea
\prod_{i<j}\psi(\theta_{ji})^{2\alpha' k_i \cdot k_j} = \left[\frac{\psi(\theta_{43})\psi(\theta_{2})}{\psi(\theta_{42})\psi(\theta_{3})}\right]^{-\alpha' s}\left[\frac{\psi(\theta_{41})\psi(\theta_{32})}{\psi(\theta_{42})\psi(\theta_{3})}\right]^{-\alpha' t}
\eea
this implies that we need the following approximations:
\bea
\left[\frac{\psi(\theta_{43})\psi(\theta_{2})}{\psi(\theta_{42})\psi(\theta_{3})}\right]^{-\alpha' s} &\sim & \exp\{-\alpha's\,\theta_{32}\,(\pi-\theta_4)(\ln \psi)''\}\nn\\
\left[\frac{\psi(\theta_{41})\psi(\theta_{32})}{\psi(\theta_{42})\psi(\theta_{3})}\right]^{-\alpha' t} &\sim & \left(\frac{\theta_{32}(\pi-\theta_4)}{\psi^2(\theta_3)}\right)^{-\alpha't}
\eea
The dominant term in $K$ for this limit is
\bea
K \sim \frac{1}{4} \ee{2}{3}\ee{1}{4} s^2
\eea 
Using the approximations above, we see that we need to compute the integral
\bea
I\equiv \int_0^{\epsilon}dx \int_0^{\epsilon}dy \, (xy)^a e^{-xyk}
\eea
in the limit when $k\to \infty$. After some algebra this becomes
\bea
I &=& k^{-a-1}\left[\ln(\epsilon^2)\int_0^{\epsilon^2k}dz \, z^a e^{-z}+\ln k \int_0^{\epsilon^2k}dz \, z^a e^{-z}-\int_0^{\epsilon^2k}dz \, z^a e^{-z}\ln z\right]\nn\\
&\sim & k^{-a-1}\left[\Gamma(1+a)\ln k -a^{-1}\Gamma'(1+a)\right]+\mathcal{O}(k^{-a-2}\ln k)
\label{estimatelargek}
\eea
Thus, the Regge limit of the amplitude is  
\bea
A_P+A_M \sim g^4 (-\alpha's)^{1+\alpha't} \Gamma(-\alpha't) \ln(-\alpha's) \Sigma(t)
\label{fullregge}
\eea
where
\bea
\Sigma(t) \equiv \alpha't \int_0^1 \frac{dq}{q} \int_0^{\pi} \left(\psi^{2\alpha't}[-\ln \psi'']^{\alpha't-1}-\psi_N^{2\alpha't}[-\ln \psi_N'']^{\alpha't-1}\right)\label{sigmanew}
\eea
and $\psi_N(\theta,q^2)=\psi(\theta,-q^2)$. This completes the calculation of the asymptotic behavior of the amplitude in the Regge limit. Notice also that the function $\Sigma(t)$ gives the one-loop correction to the open string Regge trajectory. This can be easily seen as follows. At tree level, Regge behavior implies that the 
amplitude is of the form $A\sim \beta(t) s^{\alpha(t)}$, with $\alpha(t)=1+\alpha' t$. 
Including one-loop corrections modifies both the Regge trajectory $\alpha(t)$ and the residue $\beta(t)$ by small corrections, say, $\delta \alpha$ and $\delta \beta$ respectively, i.e.,
\bea
(\beta(t)+\delta \beta) s^{\alpha(t)+\delta \alpha} \sim \beta s^{\alpha(t)} + \beta s^{\alpha(t)}\delta \alpha  \log s + \delta \beta s^{\alpha(t)}\, ,
\eea
Thus, the new trajectory is $\alpha(t)_{\rm new}= 1+\alpha't + \delta \alpha$ is captured by the term 
containing the $\log s$ factor above. Given that we are also interested in recovering the Regge behavior from the hard-scattering limit, we need to extract the large $t$ limit of $\Sigma(t)$. In order to do so we 
re-write the integral \eqref{sigmanew} as
\bea
\Sigma(t) = \alpha't \int_0^1 \frac{dq}{q} \int_0^{\pi} \left(e^{\alpha't\ln(-\psi^2[\ln \psi]'')}[-\ln \psi]''^{-1}-e^{\alpha't\ln(-\psi_N^2[\ln \psi_N]'')}[-\ln \psi_N]''^{-1}\right)
\eea
from where we see that its leading behavior at large $t$ is given by the critical points of $\ln(-\psi^2[\ln \psi]'')$ and $\ln(-\psi_N^2[\ln \psi_N]'')$. Notice that now we only have a two-dimensional integration region, for which the critical points should be easier to analyze in principle. The leading contribution comes from the $q\sim 0$ region, thus we also need the approximations:
\bea
[-\ln \psi]'' \sim [-\ln \psi_N]'' &\sim&  \csc^2\theta\nn\\
\ln(-\psi^2[\ln \psi]'')\sim -\ln(-\psi_N^2[\ln \psi_N]'') &\sim & 16 q^2 \sin^4\theta
\eea
Notice that the regions $\theta\sim0,\pi$ also produce important contributions to the integral for large $t$ and need to be analyzed separately. For this purpose we would need the corresponding asymptotic expressions for the functions $\psi$ and $\psi_N$ and to integrate over the full range $0<q<1$ (figure 
$\ref{fig:planarloop}$ shows the diagram corresponding to the planar amplitude at fixed $q$). We will come back to this point at the end of this section and we will find that these regions produce subleading behavior with respect to the contribution coming from $q\sim 0$. 

For small $q$, $\Sigma(t)$ becomes
\bea
\Sigma(t) \sim i\alpha't \int_0^{\epsilon} \frac{dq}{q} \int_0^{\pi} d\theta \, \sin^2\theta\left(e^{i 16 \, q^2 \sin^4\!\theta \, \alpha't}-e^{-i 16 \, q^2 \sin^4\!\theta \, \alpha't}\right)
\label{smallq}
\eea
Note that we have also defined the integral above by analytical continuation ($t\to i t$) as in \cite{GrossManes}. Therefore, we wish to obtain the large $|t|$ behavior of the expression
\bea
\Sigma(t) &\sim& i\, \alpha't \int_{\delta}^{\pi-\delta} d\theta \, \sin^2\theta \int_0^{\epsilon} \frac{dq}{q} \left(e^{i t a q^2}-e^{-i t a q^2}\right)\nn\\
\eea
for fixed $\epsilon$ with $a=16 \sin^4\theta$. We have also introduced the cutoff $\delta$ to stress the fact that we need to examine the contributions from the regions where $\theta \sim 0,\pi$ separately. Performing the change $a t q^2 \equiv u$ we have
\bea
\Sigma(t) &\sim& i\,\alpha't \int_{\delta}^{\pi-\delta} \sin^2\theta \, d\theta \, i \int_0^{\epsilon^2 t a} \frac{du}{u} \sin u
\eea
Since $\epsilon$ is small but fixed we can take the upper limit of the $u$ integral to be $\infty$ in the $|t|\to\infty$ limit. Also, in this limit, the $\theta$ dependence in the integration over $u$ disappears which allows us to send $\delta$ to zero, thus
\bea
\Sigma(t) &\sim&  -\alpha't \int_{0}^{\pi} \sin^2\theta \, d\theta  \int_0^{\infty} \frac{du}{u} \sin u = -\alpha't \, \frac{\pi^2}{4}
\eea
Therefore, continuing back to $t\to - it$ we have
\bea
\Sigma(t) &\sim& i \alpha't \quad \mbox{as} \quad t\to -\infty
\label{sigmalarget}
\eea
Finally, as $t\to -\infty$, combining equations \eqref{fullregge} and \eqref{sigmalarget} yields
\bea
A_P+A_M &\sim & i(-\alpha's)^{1+\alpha't} \Gamma(-\alpha't) \ln(-\alpha's) \alpha't\nn\\
&=& i(-\alpha's)^{1+\alpha't} \Gamma(1-\alpha't) \ln(-\alpha's)
\eea
We could use Stirling's approximation $\Gamma(1-\alpha't)\sim \sqrt{2\pi}(-\alpha't)^{1/2-\alpha't}e^{\alpha't}$ valid for $-\alpha't\gg 1$, which yields
\bea
A_P+A_M \sim i(-\alpha's)^{1+\alpha't} (-\alpha't)^{1/2-\alpha't} e^{\alpha't} \ln(-\alpha's)
\label{reggelarget2}
\eea
To conclude, we take a moment to analyze the regions where $\theta\sim0,\pi$ which are also important as $|t|$ becomes large. Using the following expression for the logarithm of $\psi$
\bea
\ln \psi(\theta) = \ln \sin\theta + 2\sum_{n=1}^{\infty} \frac{1}{m} 
\frac{q^{2m}}{1-q^{2m}}(1-\cos2m\theta)
\eea
one can see that
\bea
\ln(-\psi^2[\ln \psi]'')\sim -\ln(-\psi_N^2[\ln \psi_N]'') &\sim & \mathcal{O}(\theta^4)
\eea
Thus, the main contribution at large $t$ comes from the region where $\theta$ is of the order of $\sim (-\alpha't)^{-1/4}$. A rough estimation from these regions gives $\Sigma(t)\sim (-\alpha't)^{-3/4}$ which is subleading with respect to the $q\sim 0$ contribution given in \eqref{sigmalarget}.
\subsection{Hard scattering at one loop}
The high-energy limit at fixed scattering angle for the one-loop amplitude was first computed by \cite{AAM} in the early days of string theory in the context of the old dual resonance models. There, the computation was done for the non-planar amplitude which had a dominant saddle point in the interior of the integration region. In \cite{GrossManes}, Gross and Ma\~nes showed that only the (2+2) non-planar amplitude (i.e. the amplitude with two particles on each boundary of the annulus) has a saddle point in the interior of the region of integration. The planar, non-orientable and the (3+1) non-planar amplitudes do not possess a dominant saddle point in the interior, but points in the boundary of the region do give sub-dominant contributions (with respect to the (2+2) non-planar) from the boundaries of the region of integration. They also showed that the leading contribution for the sum of the planar and non-orientable diagrams comes from the region where $q \sim 0$ and the cross ratio $x\equiv \frac{\sin \theta_2 \sin\theta_{43}}{\sin\theta_{42}\sin\theta_3}$ is approximately $(1+t/s)^{-1}$ in equation \eqref{full}. We begin this section by re-calculating the leading behavior known in the literature using the saddle point approximation for the cross-ratio although using a different set of integration variables \cite{NS} where $\theta_2 \to x$, $\theta_3\to r\equiv \sin\theta_{43}/ \sin\theta_3$. Starting from equations \eqref{full} and \eqref{fulldef} the relevant factor in the integrand in this limit is
\bea
\prod_{i<j}\psi^{2\alpha' k_i\cdot k_j} = \exp\{-\alpha's V_{\lambda}\}
\eea
where
\bea
 V_{\lambda} &\equiv& \ln x -\lambda \ln(1-x) + 2\sum_{n=1}^{\infty} \frac{1}{n} 
\frac{q^{2n}}{1-q^{2n}}(S_n-\lambda T_n)\\
x&\equiv&\frac{\sin \theta_2 \sin\theta_{32}}{\sin\theta_{42}\sin\theta_3}\\
S_n &\equiv& 2\cos n(\theta_2-\theta_{43})  \left[\cos n(\theta_{42}+\theta_3) 
-\cos n (\theta_2+\theta_{43}) \right]\nn\\
T_n &\equiv& 2\cos n(\theta_{42}+\theta_{3})  \left[\cos n(\theta_{2}-\theta_{43}) -\cos n (\theta_2+\theta_{43}) \right]
\eea
and $\lambda =-t/s$. Expanding the function $V_{\lambda}$ about the critical region mentioned above yields
\bea
e^{-\alpha's V_{\lambda}} &\approx& e^{-\mathcal{E}_0}\, e^{-\alpha's[\frac{(1-\lambda)^3}{2\lambda}(x-x_c)^2\pm 2q^2(S_1-\lambda T_1)]}
\eea
where 
\bea
\mathcal{E}_0 &\equiv & \alpha'|s|[\lambda \ln(-\lambda)+(1-\lambda)\ln(1-\lambda)]\nn\\
&=&\alpha's\ln(-\alpha's)
+\alpha't\ln(-\alpha't)+\alpha'u\ln(\alpha'u)
\eea
In the $|s|\to \infty$ limit, the integration over $x$ can be approximated by a gaussian giving
\bea
\int_{-\infty}^{\infty} dx \, e^{-\alpha's \frac{(1-\lambda)^3}{2\lambda}(x-x_c)^2} \sim \sqrt{\frac{-2\pi\lambda}{(1-\lambda)^3}} (-\alpha's)^{-1/2}
\eea
The integral over $q$ is dominated by the small $q$ region which, after analytic continuation to $s\to i s$ behaves as
\bea
\int_0^{\epsilon} \frac{dq}{q} \, \left(e^{\,2i\alpha's\,q^2(S_1-\lambda T_1)}- e^{-2i\alpha's\,q^2(S_1-\lambda T_1)}\right) \sim \frac{i\pi}{2}
\label{qint}
\eea
result which we already encountered in \eqref{smallq}. All in all, for the coefficient of $\ee{1}{4}\,\ee{2}{3}$, we obtain:
\bea
A_P+A_M &\sim& su \, e^{-\mathcal{E}_0}\sqrt{\frac{-2\pi\lambda}{(1-\lambda)^3}} (-\alpha's)^{-1/2} F(\lambda)\nn\\
&\sim& s^2(1+t/s) \, e^{-\mathcal{E}_0} (-\alpha't)^{1/2}(-\alpha's)^{-1/2}(1+t/s)^{-3/2}(-\alpha's)^{-1/2}F(\lambda)\nn\\
&\sim& (-\alpha's)^{3/2}\, e^{-\mathcal{E}_0} (-\lambda)^{1/2}(1-\lambda)^{-1/2}F(\lambda)
\eea
which shows the usual exponential suppression $e^{-\mathcal{E}_0}$ factor and where the function $F(\lambda)$ is given by
\bea
F(\lambda) = \int_0^{\infty}dr \int_0^{\pi} d\theta \, \frac{r\sin^2\theta}{(r^2+2r\cos\theta+1)(r^2(1-\lambda)^2+2r(1-\lambda)\cos\theta+1)}
\label{Flambda}
\eea
An few remarks are important to note about this integral. Since $-\infty<\lambda<0$ it is convergent in this entire range but it diverges for $\lambda=0$. As $\lambda$ gets closer to zero, the integral becomes larger and larger and we need to estimate how it diverges in order to extract the correct small $\lambda$ behavior. As we will show in the next section, we have that
\bea
F(\lambda) \sim -2\ln(-\lambda)+2\ln(1-\lambda) \sim 2\ln(-\alpha's) \qquad \mbox{for $s\gg t$}
\label{Flambdasmall}
\eea 
which provides the logarithm that appears in the Regge limit of the amplitude in \eqref{reggelarget2}.
Writing the exponential factor as
\bea
e^{-\mathcal{E}_0} = (-\alpha's)^{\alpha't} (-\alpha't)^{-\alpha't} (1+t/s)^{\alpha's+\alpha't}
\eea
we have
\bea
A_P+A_M &\sim&  i (-\alpha's)^{1+\alpha't} (-\alpha't)^{1/2-\alpha't}(1+t/s)^{\alpha's+\alpha't-1/2}
F(\lambda)
\label{finalfa}
\eea
which completes the hard scattering limit of the one-loop amplitude.  
\section{Recovery of the Regge limit}
The high-energy behavior at fixed angle given in Eq. \eqref{finalfa} uses a gaussian approximation around the dominant saddle point given by $x_{c} = (1-\lambda)^{-1}$. We will now calculate this limit using a different method which does not require the gaussian approximation but instead we will compute the integral over the $x$ variable in an exact closed form. However, we still need to approximate the exponent for small $q$ but this is not too serious since this is the only place in the $q$ integration where there is dominant critical point \cite{GrossManes}. One could regard the calculation we perform in this section as a computation of the gaussian approximation including all the possible fluctuations around the saddle. This allows us to bypass the issue of computing the contributions coming any other region in the $\theta_k$ integrations since we will be computing this triple integral in exact form. Starting from \eqref{full}, we obtain
\bea
\prod_{i<j}\psi(\theta_{ji})^{2\alpha'k_i\cdot k_j} &=& 
e^{-\alpha's V_{\lambda}} \approx e^{-\alpha's\left[\ln x-\lambda \ln(1-x)+2q^2(S_1-\lambda T_1)\right]}\nn\\
& \approx & x^{-\alpha's}(1-x)^{-\alpha't} e^{-2\,\alpha's \,q^2(S_1-\lambda T_1)}
\eea
Notice that this time we are not expanding the function $\ln x-\lambda \ln(1-x)$ about the saddle point $x_c$. The small $q$ contribution to the total amplitude can be written as
\bea
A_P+A_N \sim \alpha'^2 su \int\prod_k d\theta_k  \,\, x^{-\alpha's}(1-x)^{-\alpha't} 
\int_0^{\epsilon} \frac{dq}{q} \left[ e^{-2\,\alpha's \,q^2(S_1-\lambda T_1)}-e^{2\,\alpha's \,q^2(S_1-\lambda T_1)}\right]
\eea
where we have included the overall $\alpha'^2 su$ coefficient coming from the coefficient of $\ee{1}{4} \ee{2}{3}$. We have already encountered the expression for the $q$ integral above with the very satisfying result that it does not depend on the coefficient of $q^2$ in the exponent, therefore it does not bring an angular dependence from the combination $S_1-\lambda T_1$ which will allow us to perform an exact evaluation of the integration over the $\theta_k$ variables. The integral
\bea
I\equiv \int_0^{\pi}d\theta_4\int_0^{\theta_4}d\theta_3\int_0^{\theta_3}d\theta_2 \,\, x^{-\alpha's}(1-x)^{-\alpha't}
\label{residue} 
\eea
was evaluated long ago by Green and Schwarz \cite{GS} in the context of proving that dilaton tadpole divergences could be absorbed in a renormalization of the Regge slope $\alpha'$. This was realized before it was recognized that this divergence is absent for the $SO(32)$ gauge group. We simply quote the answer here
\bea
I = \int\prod_k d\theta_k  \,\, x^{-\alpha's}(1-x)^{-\alpha't} = \gamma \frac{1}{\alpha'}\frac{\partial}{\partial \alpha'} \left[\alpha'\frac{\Gamma(-\alpha's)\Gamma(-\alpha't)}{\Gamma(1-\alpha's-\alpha't)} \right]
\label{GSInt}
\eea
where $\gamma$ is a numerical constant.
Using this and the result for the integral over $q$ given in eq. \eqref{qint} we have
\bea
A_P+A_N \sim  i \alpha'^2 su \frac{1}{\alpha'}\frac{\partial}{\partial \alpha'} \left[\alpha'^2\frac{\Gamma(-\alpha's)\Gamma(-\alpha't)}{\Gamma(1-\alpha's-\alpha't)} \right]
\eea
where we have omited the numerical coefficient $\gamma$ for simplicity.
We can now take the limit $s,t\to-\infty$ holding $t/s$ fixed directly inside the brackets to obtain
\bea
A_P+A_N \!\!\!&\sim&\!\!\! i \alpha'^2 su \frac{1}{\alpha'}\frac{\partial}{\partial \alpha'} 
\left[\alpha'^2 (-\alpha's)^{-1+\alpha't}(-\alpha't)^{-1/2-\alpha't}(1+t/s)^{-1/2+\alpha's+\alpha't}\right]\nn\\
\!\!\!&\sim&\!\!\!  i(-\alpha's)^{1/2}  (-\lambda)^{-1/2-\alpha't}(1-\lambda)^{1/2+\alpha's+\alpha't}\left[1+2\alpha's\left(\lambda\ln(-\lambda)
+(1-\lambda)\ln(1-\lambda)\right)\right]\nn\\
{}
\eea
Taking again $\alpha's\gg 1$, we end up with
\bea
A_P+A_N\!\!\!&\sim&\!\!\! i (-\alpha's)^{3/2} (-\lambda)^{-1/2}(1-\lambda)^{1/2}e^{\alpha's[\lambda\ln(-\lambda)
+(1-\lambda)\ln(1-\lambda)]}\left[\lambda\ln(-\lambda)
+(1-\lambda)\ln(1-\lambda)\right]\nn
\eea
To recover the Regge behavior we take $s\gg t$ above. The exponential becomes
\bea
e^{\alpha's[\lambda\ln(-\lambda)
+(1-\lambda)\ln(1-\lambda)]} = (-\lambda)^{-\alpha't}(1-\lambda)^{\alpha's+\alpha't}\sim (-\alpha's)^{\alpha't}(-\alpha't)^{-\alpha't}e^{\alpha't}
\eea
and the last factor becomes
\bea
\left[\lambda\ln(-\lambda)
+(1-\lambda)\ln(1-\lambda)\right] \sim \lambda \ln(-\lambda)&=&-t/s\,[\ln(-\alpha't)-\ln(-\alpha's)]\nn\\
&\sim&(-\alpha't)(-\alpha's)^{-1}\ln(-\alpha's)
\eea
Therefore, the Regge limit at high $t$ is
\bea
A_P+A_N\!\!\!&\sim&\!\!\! i (-\alpha's)^{3/2} (-\lambda)^{-1/2} (-\alpha's)^{\alpha't}(-\alpha't)^{-\alpha't}e^{\alpha't} (-\alpha't)(-\alpha's)^{-1}\ln(-\alpha's)\nn\\
\!\!\!&\sim&\!\!\!  \,i(-\alpha's)^{1+\alpha't} (-\alpha't)^{1/2-\alpha't}e^{\alpha't}\ln(-\alpha's)
\label{hardsmallangle}
\eea
which is exactly the result we found in \eqref{reggelarget2}.\\
We finish this section by showing that the result in \eqref{hardsmallangle} can also be obtained from the approximate expression in \eqref{finalfa} by analyzing the small $\lambda$ behavior of $F(\lambda)$ as anticipated in \eqref{Flambdasmall}. We believe it is instructive to do this because we are also interested in the small $\lambda$ behavior of the hard scattering limit of the type 0 model in the context of \cite{RojasThorn,Thorn:Digital,Thorn:Summing} where we cannot afford the luxury of having an exact expression for the coefficient of the exponential falloff. For convenience we write this integral here again
\bea
F(\lambda) = \int_0^{\infty}dr \int_0^{\pi} d\theta \, \frac{r\sin^2\theta}{(r^2+2r\cos\theta+1)(r^2(1-\lambda)^2+2r(1-\lambda)\cos\theta+1)}
\label{Flambda2}
\eea
As mentioned above, $F(\lambda)$ diverges as $\lambda \to 0$. The only singular region in this limit is $\theta \sim \pi$ and $r\sim 1$. It is straightforward to see this by recalling that, in terms of the cross ratio $x$, the the dominant saddle point is given by $x_c=(1-\lambda)^{-1}$. This perfectly matches with the fact that the Regge behavior of the amplitude is obtained from the region $\theta_2\sim\theta_3$, $\theta_4\sim \pi$ since $x \sim \theta_{32}(\pi-\theta_4)$ which gives the leading behavior \cite{RojasThorn,OPE}. Thus, the Regge limit occurs when $x\to 1$. Therefore in the small scattering angle limit, the integral above is singular where $\theta\sim\pi$, $r\sim x$, thus
\bea
F(\lambda) &\sim& \int_{x-\delta}^{x+\delta}dr \int_{\pi-\epsilon}^{\pi} d\theta \, \frac{(\pi-\theta)^2}{((x-1)^2+x(\pi-\theta)^2)((r/x-1)^2+(\pi-\theta)^2)}\nn\\
&\sim& 2 \int_0^{\epsilon} \frac{\theta}{(x-1)^2+x\theta^2}=-2\ln|1-x|+\ln((1-x)^2+\epsilon^2)
\label{Flambda2}
\eea
Therefore, as $\lambda\to0$  for fixed $\epsilon$, we have
\bea
F(\lambda) &\sim&  -2 \left(\ln(-\lambda)-\ln(1-\lambda)\right) \sim 2 \ln (-\alpha's)
\eea
as anticipated in \eqref{Flambdasmall}.
\section{Conclusions}

By studying the hard scattering limit of the sum of the one-loop planar and non-orientable diagrams of type I superstrings in flat 
spacetime, 
we found the exact dependence in the scattering angle that multiplies the known exponential suppression at high energies. This avoids the issue of having to estimate the contributions from flat directions in the angular integrals and the fluctuations around the saddle point, since we have at our disposal an exact result for the triple integral over the angular variables (i.e. the integral over the moduli representing the positions of the vertex operators) in a closed form. This allowed us to compare both, the hard scattering and Regge regimes of the amplitude, since they should coincide in the limit of high-momentum transfer of the latter regime. We indeed confirmed that this matching occurs by making use of the closed form of the angular integrals given in \eqref{GSInt}. As a check, we were also able to obtain this result from the approximate expression \eqref{finalfa} by analyzing the behavior of the integral $F(\lambda)$ in \eqref{Flambda} as $\lambda \to 0$.  \\
An immediate extension of this work would be to allow the open strings to be attached to smaller dimensional D$p$-brane (here we considered the case of a space-filling D-branes) where the small $q$ behavior can be analyzed separately for the planar and non-orientable diagrams since the amplitudes are finite as long as $p<8$. It would also be interesting to check if our methods could be also applied to the situation studied in \cite{Bachas}, where the authors analyzed the case where the two colliding open strings lived on different D-branes separated by a fixed distance.

\subsubsection*{Acknowledgements}
I wish to thank professor Charles Thorn for many valuable discussions and constant encouragement. I also thank Juan I. Jottar for comments on the manuscript, Andr\'e Neveu and Juan L. Ma\~nes for correspondence, and to Chrysostomos Kalousios, Jay Perez, Christoph Sachse and Heywood Tam for useful discussions and support. This research was supported in part by the Department of Energy under Grant No. DE-FG02-97ER-41029.

\end{document}